\documentclass{article}
\usepackage{spconf,amsmath,graphicx,multirow,url,cite}

\title{AN IMAGE IDENTIFICATION SCHEME OF ENCRYPTED JPEG IMAGES FOR PRIVACY-PRESERVING PHOTO SHARING SERVICES}
\name{Kenta Iida and Hitoshi Kiya}
\address{Tokyo Metropolitan University, Asahigaoka, Hino-shi, Tokyo, Japan}
\begin{document}
\ninept
\maketitle
%
%
\begin{abstract}
We propose an image identification scheme for double-compressed encrypted JPEG images that aims to identify encrypted JPEG images that are generated from an original JPEG image. 
To store images without any visual sensitive information on photo sharing services, encrypted JPEG images are generated by using a block-scrambling-based encryption method that has been proposed for Encryption-then-Compression systems with JPEG compression. 
In addition, feature vectors robust against JPEG compression are extracted from encrypted JPEG images.
The use of the image encryption and feature vectors allows us to identify encrypted images recompressed multiple times.
Moreover, the proposed scheme is designed to identify images re-encrypted with different keys.
The results of a simulation show that the identification performance of the scheme is high even when images are recompressed and re-encrypted.
\end{abstract}
\begin{keywords}
image identification, JPEG, Encryption-then-Compression system, privacy-preserving
\end{keywords}
\section{Introduction}
\label{sec:intro}
\noindent With the rapid growth of social networking services (SNSs) and cloud computing, photo sharing via various services has greatly increased. Generally, images are uploaded and stored in a compressed form to reduce the amount of data.
In the uploading process for these SNSs, it is known that service providers employ manipulation, such as recompression.
In addition, most of the content includes sensitive information, such as personal data and copyrights.
Thus, it is required that images on photo-sharing services be prevented from leakage and unauthorized use by service providers.

For privacy-preserving photo sharing on these services, three requirements need to be satisfied: 1) protection of visual information, 2) tolerance for recompression after encryption, and 3) identification of encrypted images.
In terms of requirement 1, a lot of studies on secure and efficient communications have been reported\cite{Intro1,Intro2,SNS1, SNS2,enc1,EtC1,EtC2,EtC3,EtC4,EtC5,tradesys1,tradesys2,tradesys3,tradesys4,tradesys5}.
To secure multimedia data, full encryption with provable security, such as RSA and AES, is the most secure option.
However, most schemes do not consider requirement 2, although requirement 3 is considered in several schemes\cite{tradesys3,tradesys4, tradesys5,EIR1,EIR2}.
As systems that satisfy both requirements 1 and 2, Encryption-then-Compression (EtC) systems have been developed\cite{EtC1,EtC2,EtC3,EtC4,EtC5}.
In this paper, we focus on a block-scrambling-based image encryption method that has been proposed for EtC systems.

Regarding requirement 3, image retrieval and the identification of encrypted images have never been considered for EtC systems, although image identification and retrieval schemes that are robust against JPEG compression have been proposed for unencrypted images\cite{dc1,dc2,ih1,ih2,itq}. However, the performance of these schemes is degraded in the case of identification between encrypted images and corresponding re-encrypted images, so, to satisfy all requirements, novel image identification methods are required.

Thus, we propose a novel image identification scheme for encrypted JPEG images compressed under various coding conditions.
Image encryption is carried out by extending the method based on block-scrambling for EtC systems\cite{EtC1,EtC2,EtC3,EtC4,EtC5}.
The extended method has steps for two-layer encryption.
Moreover, the feature vector used for identification is designed for images encrypted by the extended method.
The use of the two-layer encryption and features allows us to identify re-encrypted images without re-calculating the features.
Simulation results show that the proposed scheme has a high identification performance, even when images are recompressed and re-encrypted.

\section{Preliminaries}
\subsection{EtC image\label{sec:etc}}
\begin{figure}[t]
\includegraphics[width=85mm]{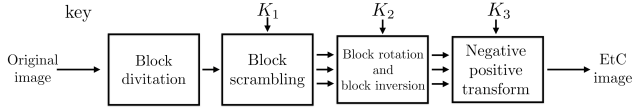}
\caption{Generation of EtC image\label{fig:enc}}
\end{figure}

\begin{figure}[t]
\begin{center}
\begin{tabular}{c}
\begin{minipage}{0.5\hsize}
  \begin{center}
   \includegraphics[width=25mm]{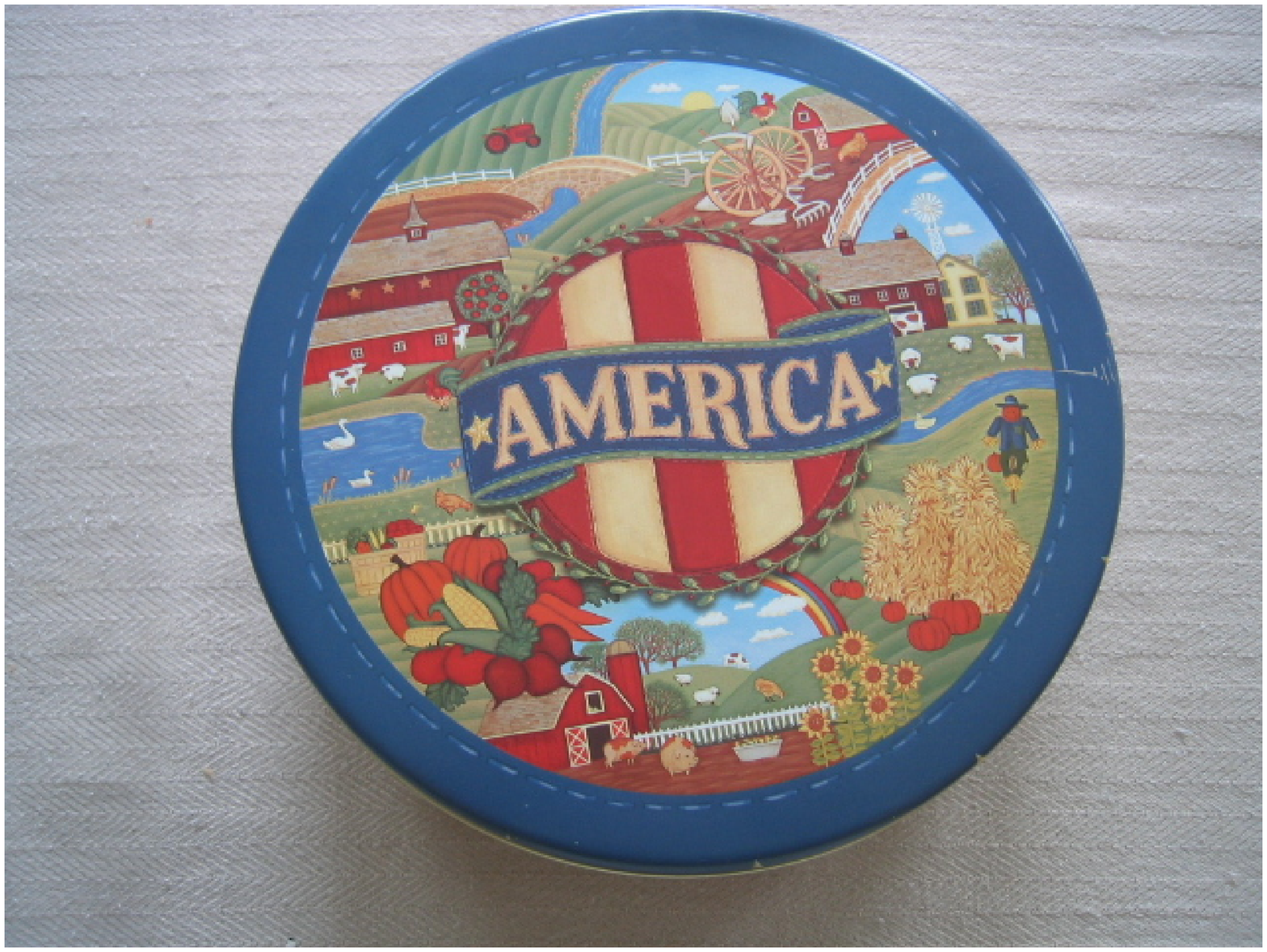}
     \hspace{16mm}(a) Original image
  \end{center}
 \end{minipage}
\begin{minipage}{0.5\hsize}
  \begin{center}
   \includegraphics[width=25mm]{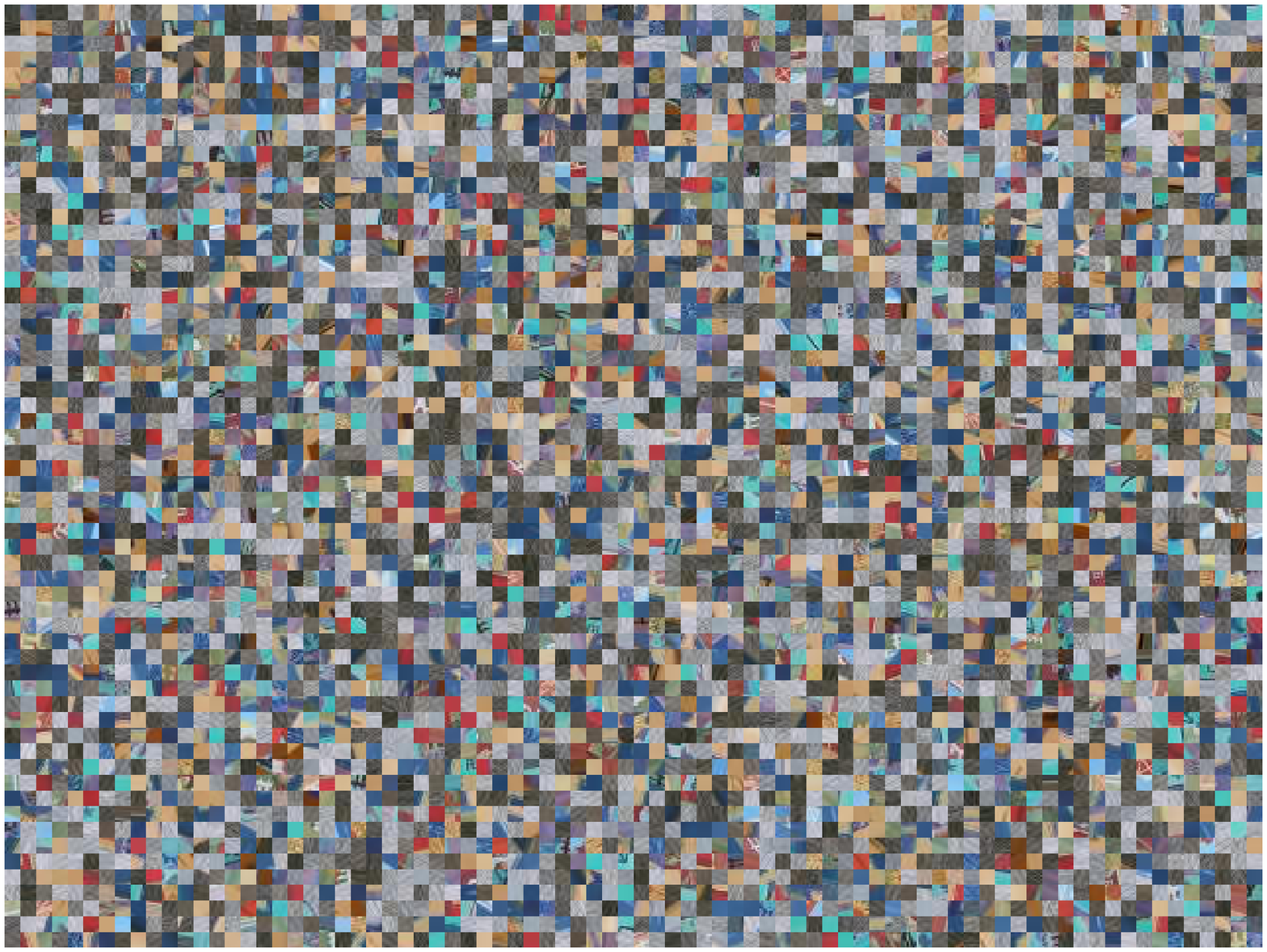}
     \hspace{16mm}(b) Encrypted image
  \end{center}
 \end{minipage}
\end{tabular}
\caption{Example of original image and encrypted image\label{fig:encgen}}
 \end{center}
\end{figure}

\noindent We focus on EtC images which have been proposed for Encryption-then-Compression (EtC) systems with JPEG compression \cite{EtC1,EtC2,EtC3,EtC4,EtC5}.
EtC images have not only almost the same compression performance as that of unencrypted images, but also enough robustness again various ciphertext-only attacks including jigsaw puzzle solver attacks\cite{EtC3}.
The procedure of generating EtC images is conducted as below (see Figs. \ref{fig:enc} and  \ref{fig:encgen})\cite{EtC2}.
\begin{itemize}
\item[1)]
Divide an image with $X \times Y$ pixels into non over-lapping $8 \times 8$ blocks. 
\item[2)]
Permute randomly $M$ divided blocks by using a random integer  secret key ${ K}_1$, where $M= \lfloor \frac{X}{8} \rfloor \times \lfloor \frac{Y}{8} \rfloor$.
\item[3)]
Rotate and invert randomly each divided block  by using a random integer  secret key ${ K}_2$.
\item[4)]
Apply a negative-positive transform to each divided block by using a random integer  secret key ${ K}_3$.
In this step, transformed pixel value at the position $(x,y)$ in a block $I_{np}(x,y)$ is computed from the  pixel value  at the same position $I(x,y)$ in a $8\times 8$ block of an image applied step from 1 to 3  to by
\begin{equation}
\label{eq:np}
I_{np}(x,y) =  \left \{
\begin{array}{l}
I(x,y),\  b=0,\\
255-I(x,y),\ b=1,
\end{array}
\right.
\end{equation}
where $b$ is a random binary value generated by ${ K}_3$ under the probability $P(b)=0.5$ and $0\leq I(x,y)$, $I_{np}(x,y) \leq 255$.
\end{itemize}
In this paper,  images encrypted by using these steps are referred to ``EtC images".

\subsection{Image Identification for JPEG images}
\noindent Let us consider a situation in which there are two or more compressed images generated under different or the same coding conditions. 
They originated from the same image and were compressed under various coding conditions.
We refer to the identification of these images as ``image identification."
Note that the aim of the image identification is not to retrieve visually similar images.

The JPEG standard is the most widely used image compression standard.
In the usual coding procedure, after color transformation from RGB space to $\mathrm{YC_{b}C_{r}}$ space and sub-samples $\mathrm{C_{b}}$ and $\mathrm{C_{r}}$, an image is divided into non-overlapping consecutive 8$\times$8-blocks.
All pixel values in each block  are shifted from [0,255] to [-128,127] by subtracting 128, and DCT is then applied to each block to obtain 8$\times$8 DCT coefficients.
After that, the DCT coefficients are quantized, and the quantized coefficients are entropy-coded.

A DC coefficient $DC$ in each block is obtained by using the following equation, where $I(x,y)$ represents a pixel value at the position $(x,y)$ in a block.
\begin{equation}
\label{eq:dc}
DC= \frac{1}{8} \sum_{x=0}^7 \sum_{y=0}^7 (I(x,y) -128)
\end{equation}
The range of the DC coefficients is [-1024,1016].
It has been reported that the features extracted from DC coefficients are effective for recompression in the conventional schemes\cite{dc1,dc2}.
Therefore, the DC coefficients are used for the identification in this paper.

\section{Proposed Scheme}
\noindent In this section, a novel two-layer image encryption method and the proposed identification scheme are explained.
The combination of them enables us to avoid the effects of not only recompression but also re-encryption.
The notations used in the following sections are shown in Tab. \ref{tab:notation}.  
\begin{table}[t!]
\caption{Notations used in this paper}
\label{tab:notation}
\centering
\scalebox{.9}{
\begin{tabular}{c|l}\hline
$O_i$  & $i$th original JPEG image\\\hline
\multirow{2}{*}{$E_i^{(j,k_0,k)}$} & $i$th  encrypted JPEG image generated from $O_i$ with\\
&  seeds $k_0$ and $k$ and compressed $j$ times (secret\\
&  keys $K_0$ and ${\bf K}$ are generated from $k_0$ and $k$ ) \\\hline
$M$ & number of $8\times 8$-blocks in image\\\hline
\multirow{2}{*}{$O_i(m)$}  & DC coefficient of $m$th block in image $O_i$\\
&  ($0 \leq m < M$)\\\hline
\multirow{2}{*}{$E_i^{(j,k_0,k)}(m)$}  & DC coefficient of luminance in $m$th block \\
& of image $E_i^{(j,k_0,k)}$ ($0 \leq m < M$)\\\hline
\end{tabular}
}
\end{table}

\subsection{Two-layer Image Encryption}
\begin{figure}[t]
\includegraphics[width=80mm]{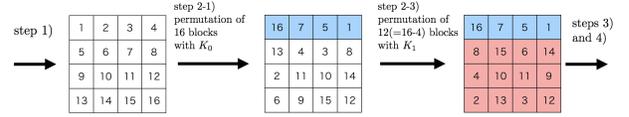}
\caption{Example of two-layer image encryption under $M=16$ and $N=4$\label{fig:twolayer}}
\end{figure}
 In the proposed scheme, the novel image encryption method, which is an extension of the method mentioned in Sec. \ref{sec:etc} for the proposed identification scheme, is conducted.
The permutation process in step 2 is divided into two layers for identification.
After dividing an image into $8\times 8$ blocks, the following processes are performed,  instead of step 2.
\begin{itemize}
\item[2-1)]
Permute randomly $M$ divided blocks using a random integer  secret key ${ K}_0$.
\item[2-2)]
Select a positive integer value $N$.
\item[2-3)]
Permute randomly the last $M-N$ blocks using a random integer  secret key ${ K}_1$ again.
\end{itemize}
After that, steps 3 and 4 are carried out.

An example of the encryption process under $M=16$ and $N=4$ is shown in Fig. \ref{fig:twolayer}.
It can be confirmed from Fig. \ref{fig:twolayer} that the first four blocks are not permuted in step 2-3.
Thus, the process in these steps does not change the positions of the first $N$ blocks under the same  $K_0$, even if  $K_1$ is changed.
This property will play an important role in the proposed identification scheme.
\subsection{Overview of Photo-sharing Services}
\begin{figure}
\includegraphics[width=85mm]{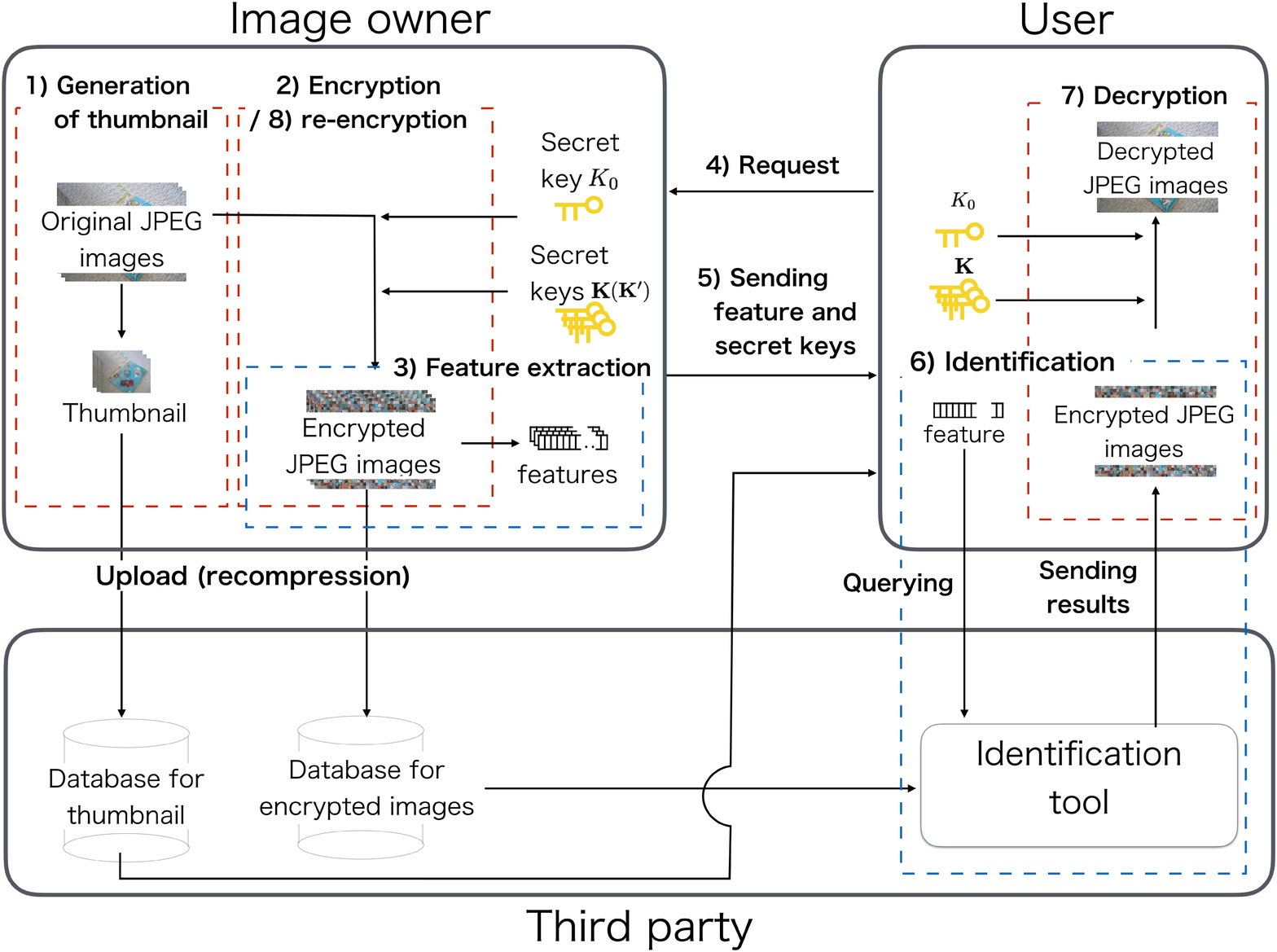}
\caption{Privacy-preserving photo-sharing\label{fig:scenario}}
\end{figure}
\noindent So that an image owner shares JPEG images on an untrusted service without publishing the sensitive information of the original quality images, a scenario of the proposed scheme is illustrated  in Fig. \ref{fig:scenario}.
\begin{itemize}
\item[1)]
An image owner generates thumbnail JPEG images from original JPEG images, and the thumbnail images are then uploaded to a third party.
\item[2)]
The image owner encrypts the original JPEG  images with secret keys $K_0$ and ${\bf K}=[K_1, K_2,K_3]$ according to the two-layer image encryption.
As shown in Fig. \ref{fig:twolayer}, the blocks of original JPEG images are permuted with the secret key $K_0$ , and encryption with ${\bf K}$ is then performed.
After that, the compressed EtC images are uploaded to the third party.
In this uploading process, these images may be recompressed.
\item[3)]
The image owner extracts features from the encrypted JPEG images.
The features are related to the uploaded thumbnail images.
\item[4)]
A user selects a thumbnail image from those stored on the third party's storage, and then sends the selected images to the image owner.
\item[5)]
The image owner sends the corresponding feature and the secret keys $K_0$ and ${\bf K}$ to the user.
\item[6)]
The third party identifies the encrypted images corresponding to the feature received from the user, and sends the identified image to the user.
\item[7)]
The user decrypts the encrypted image with $K_0$ and ${\bf K}$.
\item[8)]
The image owner re-encrypts the identified image with $K_0$ and ${\bf K'}=[K'_1, K'_2,K'_3]$ where $K_1'$, $K_2'$ and $K_3'$ are different from $K_1$, $K_2$ and $K_3$.
\end{itemize}
\vspace{-0.5\baselineskip}
Thumbnail images are not tied to corresponding encrypted images on the services because of the following two reasons.
One is to prevent jigsaw solver attacks done by using  visual information of a thumbnail image, which a third party use to  decrypt the corresponding encrypted image.
The other is to prevent the unauthorized  use of the user data. 
A third party makes it possible to collect the data on users, such as their preferences,  from a thumbnail image corresponding to an encrypted image.   
Therefore, identification using encrypted images without any visual sensitive information is required for the privacy-preserving communications.

It is known that service providers usually employ manipulation such as recompression to uploaded images.
Therefore, recompression is assumed in  steps 1 and 2 of this scenario.

The feature used in the proposed identification scheme is designed to identify images encrypted with the different key sets ${\bf K}$ and ${\bf K'}$ under the same ${K_0}$. 
Thus, recalculation of features is not needed, even when images are re-encrypted.


\subsection{Proposed Identification Scheme}
\noindent In the proposed identification scheme, the features extracted  from the first $N$ DC coefficients of Y component in the encrypted images are used.
The use of these features allow us to robustly identify images against recompression and re-encryption.
Here, the feature extraction and the identification processes are explained.
 
\subsubsection*{A. Feature extraction process}
\noindent In order to extract the feature vector of  $E_i^{(1,k_0,k)}$, the following process is performed.
\begin{itemize}
\item[(a)]
Set $N$.
\item[(b)]
Set $n:=0$.
\item[(c)]
Extract the feature vector $v_{E_i^{(1,k_0,k)}}$ from the $n$th DC coefficient of Y component in $E_i^{(1,k_0,k)}$ as below.
\begin{equation}
v_{E_i^{(1,k_0,k)}}(n) = |{E_i^{(1,k_0,k)}}(n)|,
\end{equation}
\vspace{0\baselineskip}
\vspace{-5mm}\item[(d)]
Set $n:=n+1$.
If $n < N$, return to step (c).
Otherwise, the image owner halts the process for ${E_i^{(1,k_0,k)}}$.
\end{itemize}
\vspace{-0.5\baselineskip}
In step (c), the feature is extracted from the absolute values of the DC coefficients of the Y component.
 It is known that block rotation and inversion in the DCT domain do not change the value of the DC coefficient in each block\cite{rotinv}. 
 In addition, from Eqs. (\ref{eq:np}) and (\ref{eq:dc}), the absolute value of a DC coefficient is not greatly changed by a negative-positive transform. 
 Thus, the use of DC coefficients allows us to avoid not only the effect of recompression but also that of encryption. 
 In the proposed scheme, when $K_0$ and $N$ are not changed, the DC coefficients that are moved to the first $N$ blocks are the same as those of an image encrypted with different key sets. i.e., ${\bf K}\neq {\bf K'}$. Therefore, it is expected that the absolute values of DC coefficients in the first $N$ blocks are close to those in the images encrypted with the different keys under the same $K_0$. 
 This allows us to identify re-encrypted images without re-calculating features.

\subsubsection*{B. Identification process}
As shown in Fig. \ref{fig:scenario}, a user obtains the feature vector corresponding to a thumbnail image $v_{E_i^{(1,k_0,k)}}$ from the  image owner, and then sends the vector to the third party.
The third party performs the following process to identify ${E_{i'}^{(j,k_0,k')}}$ with $E_i^{(1,k_0,k)}$, after extracting ${ v}_{E_{i'}^{(j,k_0,k')}}$ from ${E_{i'}^{(j,k_0,k')}}$.
\begin{itemize}
\item[(a)]
Set $N$ and $d$, where $d$ is a parameter that determines the acceptance error.
\item[(b)]
Set $n:=0$ and $i'=0$.
\item[(c)]
If $|v_{E_i^{(1,k_0,k)}}(n)-v_{E_{i'}^{(j,k_0,k')}}(n)|\leq d$, proceed to step (d).
Otherwise, the third party judges that $E_{i'}^{(j,k_0,k')}$ is not generated from $O_i$ and proceed to step (e).
\item[(d)]
Set $n:=n+1$.
If $n < N$, return to step (c).
Otherwise, the third party judges that $E_{i'}^{(j,k_0,k')}$ is generated from the same original image as that of $E_i^{(1,k_0,k)}$, i.e., $O_i$ and the process for $E_i^{(1,k_0,k)}$ is halted.
\item[(e)]
Set $i':=i'+1$ and $n:=0$.
If $i'$ is not equal to the number of images stored in the database of the third party, return to step (c).
Otherwise, third party judges that there is no image corresponding to $E_i^{(1,k_0,k)}$.
\end{itemize}
The distance between the absolute values of the features is used for identification under the acceptance error $d$ in step (c).
As well as the conventional schemes, this value was experimentally determined in a pre-experiment.

\section{Simulation\label{sec:sim}}
\begin{figure}[t]
\begin{center}
\begin{tabular}{c}
\begin{minipage}{0.25\hsize}
  \begin{center}
   \includegraphics[width=20mm]{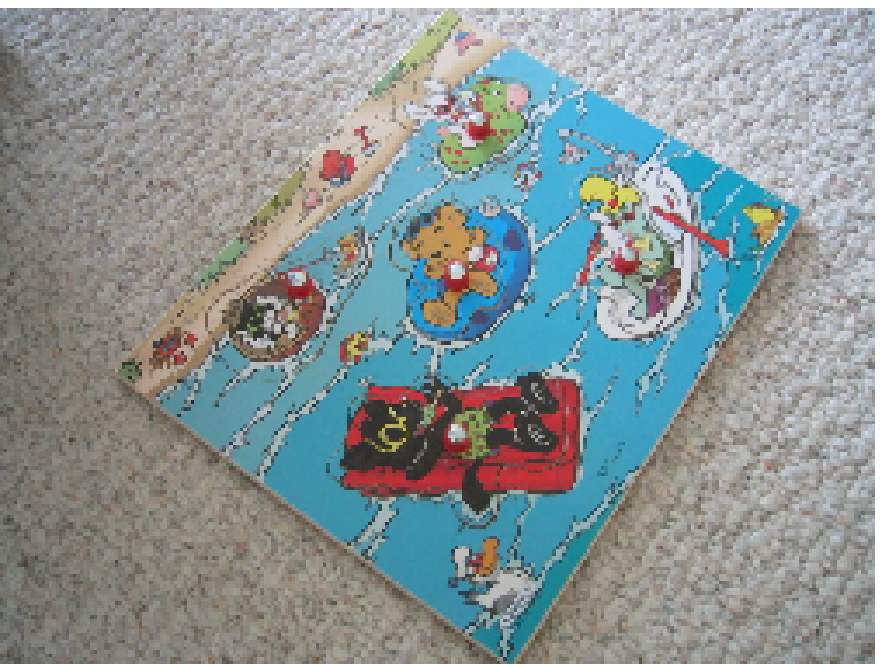}
        \hspace{16mm}(a) No.000
  \end{center}
 \end{minipage}
\begin{minipage}{0.25\hsize}
  \begin{center}
   \includegraphics[width=20mm]{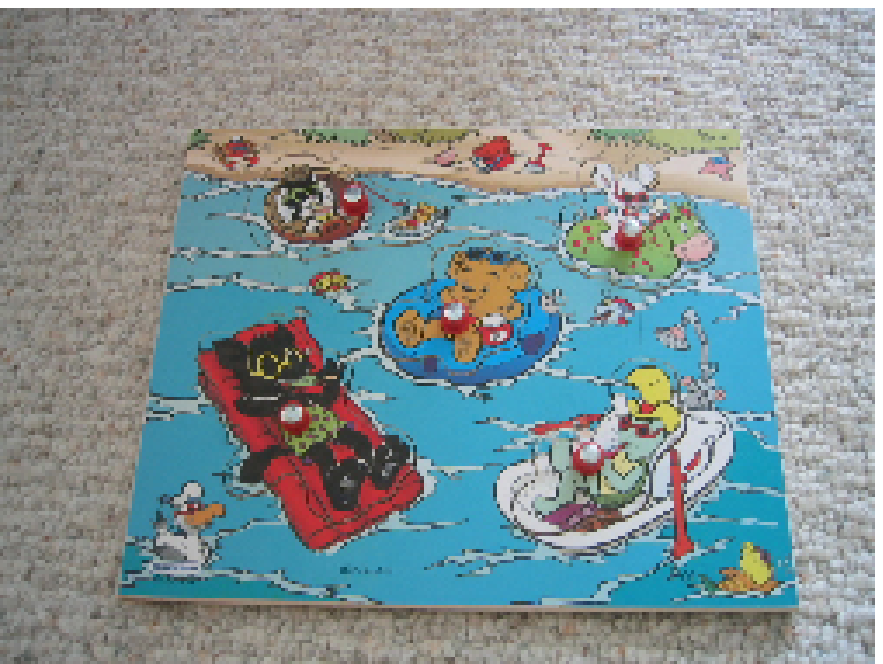}
        \hspace{16mm}(b) No.001
  \end{center}
 \end{minipage}
 \begin{minipage}{0.25\hsize}
  \begin{center}
   \includegraphics[width=20mm]{UK005.eps}
           \hspace{16mm}(c) No.004
  \end{center}
 \end{minipage}
 \begin{minipage}{0.25\hsize}
  \begin{center}
   \includegraphics[width=20mm]{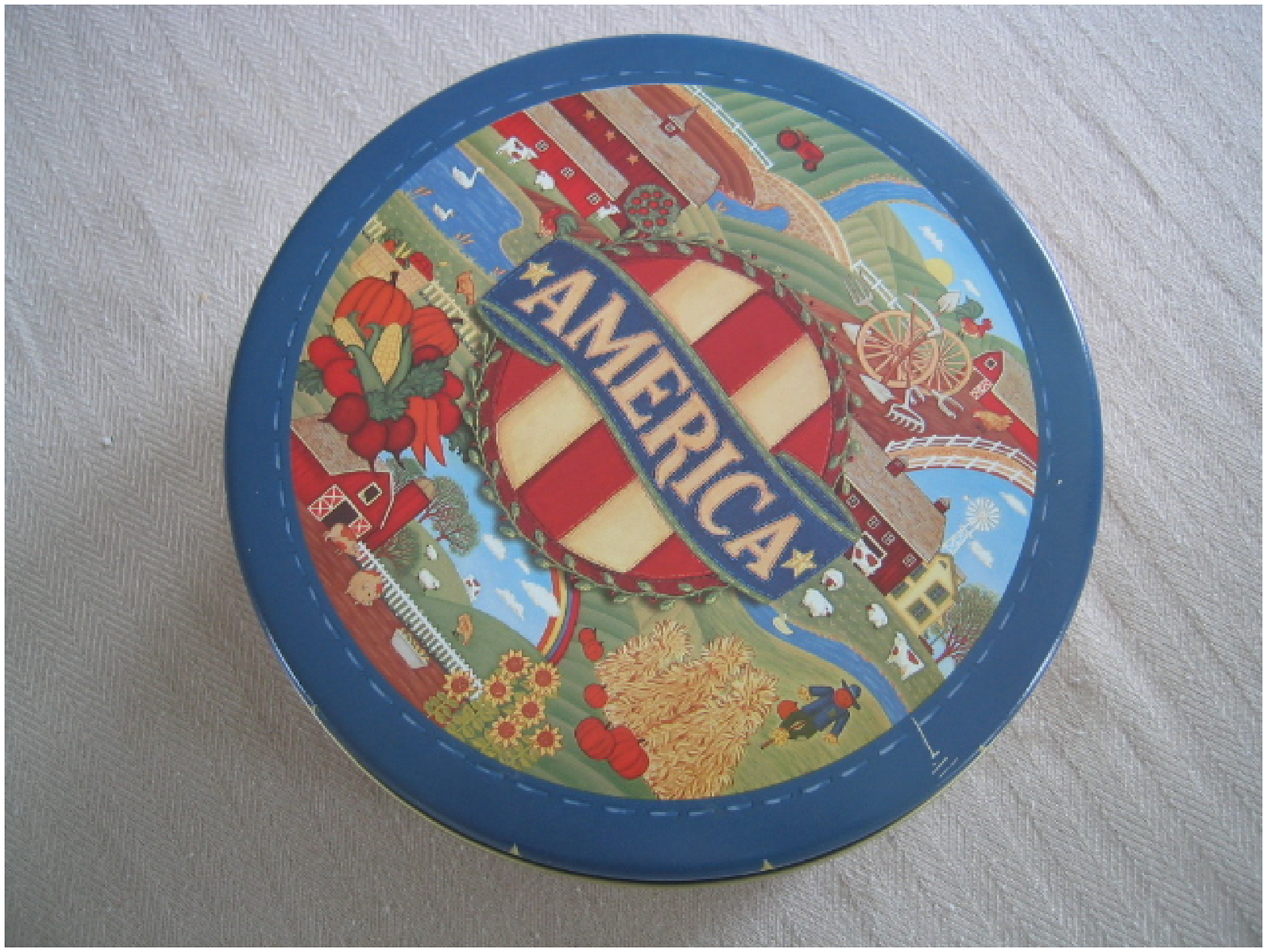}
              \hspace{16mm}(d) No.005
  \end{center}
 \end{minipage}
\end{tabular}
\caption{Examples of images in UKbench \label{fig:exuk}}
 \end{center}
\end{figure}

\noindent A number of simulations were conducted to evaluate the performance of the proposed identification scheme.
We used images (size of $480\times640$) in UKbench dataset \cite{uk} and an encoder and a decoder from the IJG (Independent JPEG Group) \cite{ijg} in the simulations.
The dataset consists of 10,200 images (4 images per 2,520 objects), and 500 images from No.000 to No.499 were chosen from  the 10,200 images (see Fig. \ref{fig:exuk}).

\subsection{Simulation Conditions} 
\noindent Table \ref{tab:condition} summarizes the conditions for generating original JPEG images and encrypted ones, where $QF_I$ indicates the quality factor used for  JPEG image $I$, $I \in \{ O_i, E_i^{(1,k_0,k)}, E_i^{(1,k_0,k')}, E_i^{(2,k_0,k)}, \\ E_i^{(2,k_0,k')}\}$.
Note that the sampling ratios for all compressions were 4:4:4.
For instance, in the case of condition (1), 500 original JPEG images were generated with $QF_{O_i}=95$ from the 500 UKbench images first.
To generate $E_i^{(1,k_0,k)}$, the 500 original JPEG images were encrypted and compressed with $QF_{E_i^{(1,k_0,k)}}=95$.
Next, these 500 single-compressed encrypted images were recompressed with $QF_{E_i^{(2,k_0,k')}}=85,80,75,70$.
As a result, 500 single-compressed and 2,000 double-compressed images encrypted with $k$ were generated under each condition.
As well as the generation of images encrypted with $k$, there were 500 single-compressed and 2,000 double-compressed JPEG images encrypted with $k'$, $QF_{E_i^{(1,k_0,k')}}=95$ and $QF_{E_i^{(2,k_0,k')}}=85,80,75,70$.
In the simulations, the identification performances between 500 single-compressed and 2,000 double-compressed images encrypted with $k$ were evaluated for each condition first.
Also, identification between 500 single-compressed images with $k$ and 2,000 double-compressed ones with $k'$ was performed.

As the parameters, $N=480$ and $d=150$ were selected. The UKbench images were divided into 4800 $8 \times 8$ blocks, i.e. $M=4800$, so that $N=480$ was selected as 10$\%$ of all blocks. To determine $d$, identification between images generated under condition (2) and $k\neq k'$ from 885 UCID images \cite{ucid} was carried out. As one of the best choices, $d=150$ was selected.

The proposed scheme was compared with four identification schemes (DC signs-based \cite{dc2}, sparse coding-based\cite{ih1}, quaternion-based \cite{ih2}, and iterative quantization (ITQ)-based ones\cite{itq}). In the schemes\cite{ih1,ih2,itq}, the Hamming distances between the hash values of the encrypted images were calculated, and images that had the smallest distance were then chosen as images generated from an original image after all images were decompressed.

\begin{table}[t!]
\caption{Condition to generate original and encrypted JPEG images, where $QF_I$ indicates the quality factor generated for the JPEG image $I$, $I \in \{ O_i, E_i^{(1,k_0,k)}, E_i^{(1,k_0,k')}, E_i^{(2,k_0,k)}, E_i^{(2,k_0,k')}\}$.}
\label{tab:condition}
\centering
\scalebox{1}{
\begin{tabular}{|c|c|c|c|c|c|c|c|c|}\hline
\multirow{2}{*}{Condition}&\multirow{2}{*}{$QF_{O_i}=$}&$QF_{E_i^{(1,k_0,k)}}=$ &$QF_{E_i^{(2,k_0,k)}}=$\\
& & $QF_{E_i^{(1,k_0,k')}}=$ &$QF_{E_i^{(2,k_0,k')}}=$\\\hline
(1) & 95 & 95 &\multirow{3}{*}{85,80,75,70}\\\cline{1-3}
(2) & 85 &  85&\\\cline{1-3}
(3) & 75 &  75&\\\hline
\end{tabular}
}
\end{table}



\subsection{Identification Performance} 
\noindent Table \ref{tab:resUK}  shows Precision value $p$ and Recall value $r$,  defined by
\begin{equation}
p = \frac{TP}{TP+FP},\ r = \frac{TP}{TP+FN},
\end{equation}
where $TP$, $FP$ and $FN$ represent the number of true positive, false positive and false negative matches respectively.
Note that $r=100[\%]$ means that there were no false negative matches, and $p=100[\%]$ means that there were no false positive matches.
\begin{table}[t!]
\caption{Identification performance for double-compressed encrypted images}
\label{tab:resUK}
\centering
\scalebox{1}{
\begin{tabular}{|c|c|c|c|c|c|c|c|c|}\hline
\multirow{2}{*}{scheme}&\multirow{2}{*}{condition}&\multicolumn{2}{|c|}{$k=k'$}&\multicolumn{2}{|c|}{$k\neq k'$}\\\cline{3-6}
& &$p$[\%]&$r$[\%]&$p$[\%]&$r$[\%]\\\hline
\multirow{3}{*}{\shortstack{proposed\\($d=150$)}}&(1)&100&100&100&100\\\cline{2-6}
&(2)&100&100&100&100\\\cline{2-6}
&(3)&100&100&100&100\\\hline
\multirow{3}{*}{\shortstack{DC sign\cite{dc2}}}&(1)&100&100&0&0\\\cline{2-6}
&(2)&100&100&0&0\\\cline{2-6}
&(3)&100&100&0&0\\\hline
\multirow{3}{*}{\shortstack{ITQ\cite{itq}}}&(1)&100&100&3.45&3.45\\\cline{2-6}
&(2)&100&100&3.25&3.25\\\cline{2-6}
&(3)&100&100&3.6&3.6\\\hline
\multirow{3}{*}{\shortstack{Sparse coding\cite{ih1}}}&(1)&99.95&100&0.09&0.15 \\\cline{2-6}
&(2)&100&100&0.33&0.55\\\cline{2-6}
&(3)&100&100&0.06&0.1\\\hline
\multirow{3}{*}{\shortstack{Quaternion\cite{ih2}}}&(1)&100&100&0.31&0.5 \\\cline{2-6}
&(2)&100&100&0.24&0.4\\\cline{2-6}
&(3)&100&100&0.56&0.95\\\hline
\end{tabular}
}
\end{table}

It was confirmed that all schemes had high identification performances under $k=k'$. However, only the proposed scheme achieved $r=100\%$ and $p=100\%$ under all conditions with $k\neq k'$; the other schemes did not. Moreover, we confirmed that the identification results with $N=480$ in Table \ref{tab:resUK} were the same as those even in the case $N \geq 160$.
\section{Conclusion}
\noindent In this paper, a novel identification scheme for encrypted images was proposed.
The image encryption is based on a block-scrambling method, and two-layer block permutation is performed in the encrypted process. 
For the identification, the feature vector used in the proposed scheme is extracted from the DC coefficients of luminance.
The use of the image encryption  method and  feature vectors allows us to avoid not only the effect of  recompression but also that of re-encryption with different keys. 
Simulation results showed the effectiveness of the proposed scheme, even when images were recompressed and re-encrypted by different keys for the second layer.


\ninept
\subsection*{Acknowledgements}
This work was partially supported by Grant-in-Aid for Scientic Research(B), No.17H03267, from the Japan Society for the Promotion Science.

\bibliographystyle{IEEEbib}
\bibliography{refv2}

\end{document}